\documentclass[conference]{IEEEtran}
\IEEEoverridecommandlockouts
\usepackage{cite}
\usepackage[numbers]{natbib}
\usepackage{amsmath,amssymb,amsfonts}
\usepackage{algorithmic}
\usepackage{graphicx}
\usepackage{textcomp}
\usepackage{xcolor}

\def\BibTeX{{\rm B\kern-.05em{\sc i\kern-.025em b}\kern-.08em
    T\kern-.1667em\lower.7ex\hbox{E}\kern-.125emX}}
\begin{document}

\title{TransVDM: Motion-Constrained Video Diffusion Model for Transparent Video Synthesis}

\author{
    \IEEEauthorblockN{Menghao Li, Zhenghao Zhang, Junchao Liao, Long Qin, Weizhi Wang}
    \IEEEauthorblockA{Alibaba Group}
}

\maketitle

\begin{abstract}
Recent developments in Video Diffusion Models (VDMs) have demonstrated remarkable capability to generate high-quality video content. Nonetheless, the potential of VDMs for creating transparent videos remains largely uncharted. In this paper, we introduce TransVDM, the first diffusion-based model specifically designed for transparent video generation. TransVDM integrates a Transparent Variational Autoencoder (TVAE) and a pretrained UNet-based VDM, along with a novel Alpha Motion Constraint Module (AMCM). The TVAE captures the alpha channel transparency of video frames and encodes it into the latent space of the VDMs, facilitating a seamless transition to transparent video diffusion models. To improve the detection of transparent areas, the AMCM integrates motion constraints from the foreground within the VDM, helping to reduce undesirable artifacts. Moreover, we curate a dataset containing 250K transparent frames for training. Experimental results demonstrate the effectiveness of our approach across various benchmarks.
\end{abstract}


\section{Introduction}
Transparent video generation involves not only creating visually compelling content but also accurately representing the alpha channels for each pixel, which indicates the degree of transparency.
This technology has broad applications across various fields, including film production and augmented reality (AR), where transparent elements are crucial for content composition and creation. However, existing video generation methods, such as Video Diffusion Models (VDMs) \cite{nichol2021glide,ho2022video,he2022latent,wang2023modelscope}, are limited to supporting only RGB channels and cannot generate transparent videos in an end-to-end manner.


In this paper, we present TransVDM, the first approach specifically designed to tackle the challenge of transparent video generation using VDMs. Unlike conventional post-processing techniques that depend on video segmentation, matting or color palette modifications, TransVDM emphasizes the direct generation of transparent videos from the outset. Inspired by LayerDiffuse~\cite{zhang2024transparent}, we initially utilized a Transparent Variational Autoencoder (TVAE) to separately encode alpha channel information as a minor perturbation within the latent space of the pretrained VDM. This process enables the VDM to incorporate alpha information within its input image, providing it the basic ability to generate transparent videos directly.
While this approach successfully facilitates transparent video generation within the existing VDM framework, it also introduces challenges, specifically that VDMs are prone to generating artifacts in transparent areas. We speculate this arises because any region in the image can be affected by the VDM, leading to pixel value changes. Consequently, this sometimes results in undesirable motion within the RGB channels of transparent areas while the corresponding alpha values transform into non-transparent values, which ultimately forms artifacts. To address this issue, we propose a lightweight Alpha Motion Constraint Module (AMCM), which incorporates the coordinates of foreground bounding boxes as motion constraints into the VDM. This integration serves to explicitly reduce the model's tendency to generate undesirable motion in transparent regions, significantly minimizing artifacts and enhancing the overall quality of the generated transparent videos.

To enable transparent video generation, we have developed a dataset consisting of 250k transparent frames for training purposes, which includes 100k high-quality transparent images and 150k transparent video frames. Our training process follows a two-stage approach: initially, we train the TVAE using the transparent images, and subsequently, we utilize the transparent videos to train the AMCM module.

Our contributions are summarized as follows: (1) We introduce TransVDM, the first diffusion model designed for transparent video generation. (2) We utilize a Transparent Variational Autoencoder to enhance the existing VDM framework and introduce a motion constraint to reduce artifacts in transparent regions. (3) We have curated a dataset consisting of 250k transparent frames for training, which can serve as a robust baseline for future research endeavors.

\section{Related Work}
The goal of current video generation is primarily to create RGB videos from a given text prompt or a combination of text and image. Early works, such as those using GANs for video generation \cite{saito2017temporal,tulyakov2018mocogan,wang2018video,li2018video} produce low-quality results. Subsequently, diffusion model based methods become popular, such as LVDM \cite{he2022latent} and ModelScopeT2V \cite{wang2023modelscope}, which transform the 2D U-Net architecture \cite{ronneberger2015u} into a 3D framework for joint training on video content. Additionally, Imagen Video \cite{ho2022imagen} synthesizes high-resolution videos from text prompts by leveraging multiple video diffusion models, while Make-A-Video \cite{singer2022make} integrates text-to-image generation with a decoupled spacetime attention mechanism. Furthermore, methods aimed at achieving fine-grained control have also gained traction, with approaches such as DragNUWA \cite{yin2023dragnuwa} and Animate-Anything \cite{dai2023fine}, which utilize motion trajectories or masks to enhance image animation. However, these methods lack support for transparency, limiting their effectiveness in certain scenarios. We address this by building the TransVDM framework, which incorporates transparency handling to improve the versatility of video generation.


\begin{figure*}[htbp]
    \centering
\includegraphics[width=0.95\textwidth]{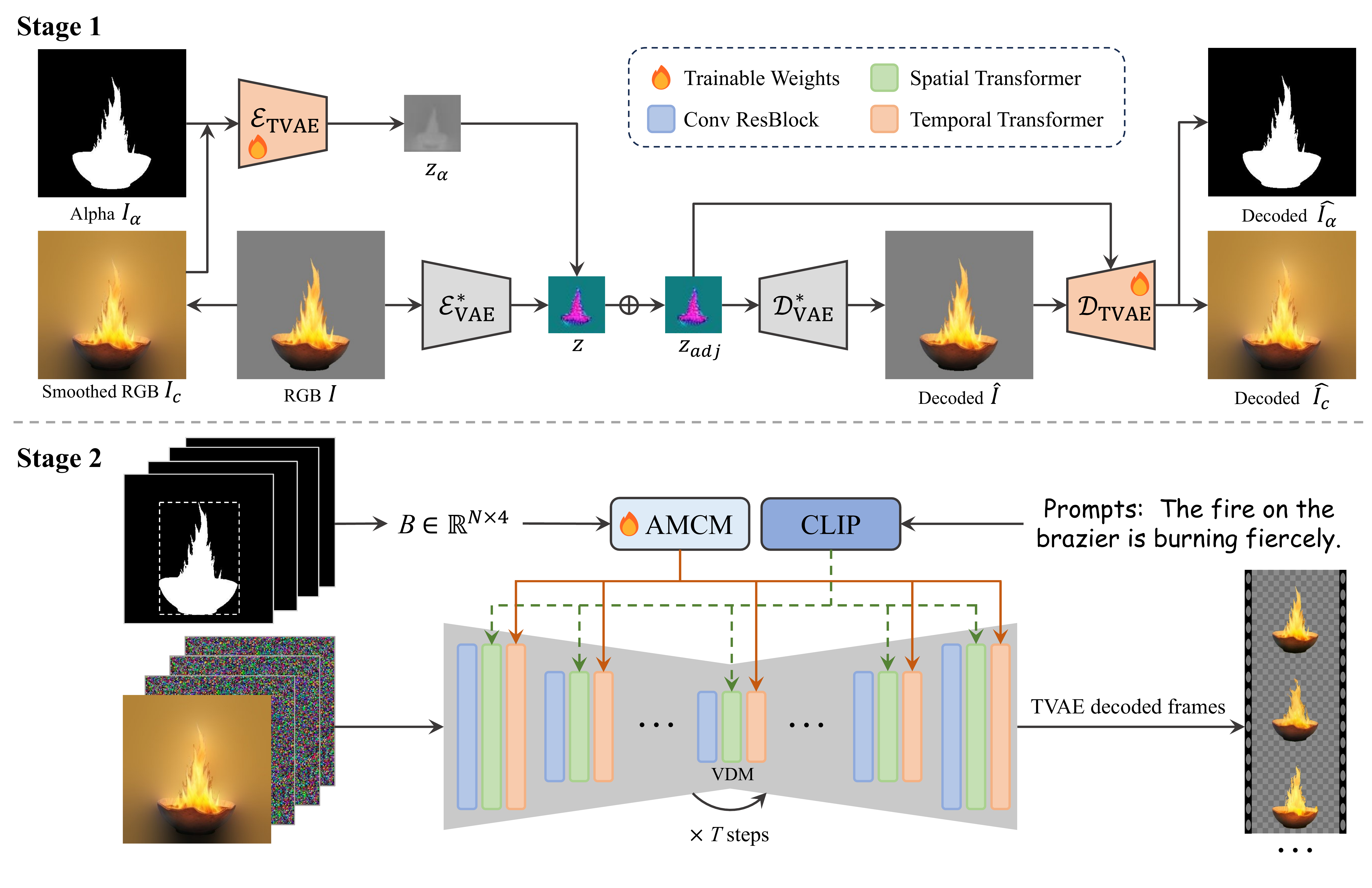}
    \caption{
    TransVDM Framework. The TVAE is trained during Stage 1, which is utilized in Stage 2 to produce adjusted noisy latents with alpha information. In Stage 1, the RGB channels receive a smoothing operation to mitigate abrupt visual artifacts that may occur when alpha edges are inaccurately predicted, allowing the transparent areas of the original image to blend more seamlessly. The Stage 2 diagram illustrates a U-Net based VDM, the concatenation of the conditioned frame and adjusted noisy latents, along with a CLIP text encoder and a simplified processing flow of the AMCM module. The AMCM module actually comprises fully connected layers and a temporal attention layer. The input features from the temporal transformer are combined with $B$, where B represents the normalized coordinates of the bounding boxes for each input frame, and processed through the AMCM module before being passed back to the temporal transformer.
    }
    \label{fig:framework}   
\end{figure*}

\section{Method}
Starting with a transparent image $I_t\in \mathbb{R}^{H \times W \times 4}$, we denote the first 3 RGB channels as $I_c \in \mathbb{R}^{H \times W \times 3}$ and the alpha channel as $I_\alpha \in \mathbb{R}^{H \times W \times 1}$. We first integrate the alpha channel $I_\alpha$ into the RGB latent space by leveraging a Transparent Variational Autoencoder (TVAE). Subsequently, we combine a pretrained image-text-to-video model which is based on AnimateAnything \cite{dai2023fine}, with our proposed Alpha Motion Constraint Module (AMCM) to finalize the TransVDM framework, as shown in Fig.\ref{fig:framework}. In section A, we provide a brief preliminary of video diffusion models, followed by an overview of TVAE in section B, a detailed explanation of AMCM in section C, and a description of our dataset in section D. 
\subsection{Preliminary}
In this section, we introduce the preliminary concepts of latent video diffusion model (LVDM) \cite{he2022latent}. Given an image $ x_0=I_c \in \mathbb{R}^{H \times W \times 3}$, the process begins by utilizing a pretrained 
vanilla Variational Autoencoder (VAE) to encode the raw image $x_0$ into the latent space $ z_0 \in \mathbb{R}^{c \times h \times w}$, which could be reconstructed by the VAE decoder. The diffusion model is optimized to predict the added noise to $z_0$ via the following objective function:
\begin{equation}
    \mathcal{L}_\epsilon = \left\|\epsilon - \epsilon_\theta(z_t, t, c)\right\|_2^2,
\end{equation}
where $c$ represents the image and text prompt,  $\epsilon_\theta$ is the noise prediction function and $z_t$ is obtained by t-step addition of Gaussian noise to $z_0$:
\begin{equation}
       z_t = \sqrt{\bar{\alpha}_t} z_0 + \sqrt{1 - \bar{\alpha}_t} \epsilon, \quad \epsilon \sim \mathcal{N}(0, I),
\end{equation}
where $t\in[0, T]$ is the time step, $\bar{\alpha}_t = \prod_{i=1}^{t} (1 - \beta_i)$ and $\beta_t$ is a coefficient with respect to t.

\subsection{Transparent Variational Autoencoder}
As shown in Fig.\ref{fig:framework} stage 1, our Transparent Variational Autoencoder (TVAE) integrates the alpha channel by separately encoding it and adding it into the RGB latent. This design ensures the alpha embedding introduces a minimal perturbation to the latent space of the vanilla VAE, avoiding changes to the overall distribution of the latent space. This enables the VAE process to transform from supporting RGB-only images to accommodating transparency information.

The TVAE consists of an encoder and a decoder. Specifically, the TVAE encoder is a simple multilayer Convolutional Neural Network (CNN) \cite{lecun1998gradient} that downscales the image by a factor of 8 in both width and height, resulting in the same dimensionality of the transparent latent $z_\alpha$ with the RGB latent $z$. The decoder mirrors that of the U-Net architecture. The TVAE is trained with two reconstruction objectives: the identity reconstruction loss and the transparency reconstruction loss. The identity loss only affects the weights of the TVAE encoder, ensuring the adjusted latent distribution remains intact, thereby enabling the vanilla VAE to accurately reconstruct RGB images. In addition, the transparency reconstruction loss is employed to jointly optimize the TVAE encoder and decoder. This yields the core TVAE procedure, which operates through the following steps:
\begin{itemize}
\item Latent Encoding: the TVAE encoder generates the latent $z_\alpha$ given an RGB image $I_c$ and the corresponding alpha channel $I_\alpha$:
\begin{equation}
    z_\alpha = \mathcal{E}_\text{{TVAE}}(I_c, I_\alpha),
\end{equation}

\item Latent Decoding: The adjusted latent $z_\text{{adj}} = z + z_\alpha$ paired with the original RGB reconstruction $\hat{I}$ is decoded back to the reconstructed transparent image:
\begin{equation}
       [\hat{I}_c, \hat{I}_\alpha] = \mathcal{D}_\text{{TVAE}}(\hat{I}, z_\text{{adj}}),
\end{equation}
where $\hat{I}_c$ and $\hat{I}_\alpha$ are the reconstructed RGB and alpha channels.

The identity loss is expressed as:
\begin{equation}
    \mathcal{L}_\text{{identity}} = ||I - \hat{I}||_2^2 = ||I - \mathcal{D}_\text{{VAE}}^*(\mathcal{E}_\text{{VAE}}^*(I)+z_{\alpha})||_2^2,
\end{equation}
where * denotes frozen models. Similarly, the transparency reconstruction loss is represented as: 
\begin{equation}
      \mathcal{L}_\text{{recon}} = ||I_c - \hat{I}_c||_2^2 + ||I_\alpha - \hat{I}_\alpha||_2^2,
\end{equation}
The final training objective is the weighted sum of two loss functions:
\begin{equation}
    \mathcal{L}_\text{{TVAE}} = \mathcal{L}_\text{{recon}} + \lambda \mathcal{L}_\text{{identity}},
\end{equation}
where $\lambda$ is a hyper-parameter.
\end{itemize}

\subsection{Alpha Motion Constraint Module}
As illustrated in Fig.\ref{fig:framework} stage 2, the backbone network is a frozen VDM. The Alpha Motion Constraint Module (AMCM) module functions as a fusion mechanism within each block of the U-Net architecture. It integrates the normalized coordinates of the bounding boxes that enclose the foreground of each alpha frame, treating them as constraint parameters that restrict the regions where motion occurs. Note that the AMCM module is comprised of several fully connected layers and a temporal attention mechanism.

Specifically, during training, the normalized coordinates of the top-left and bottom-right corners of the bounding boxes are utilized as the constraint parameters. During inference, the constraint parameters default to the values derived from the given image, meaning that the these parameters are the same for each frame. The AMCM is designed as a lightweight yet effective component. It accepts a sequence of normalized coordinates  $B \in \mathbb{R}^{N \times 4}$, where $N$ is the number of frames in the input video. The normalized coordinates are defined as:
\begin{equation}
    B = \{(x_{\text{min}}^i, y_{\text{min}}^i, x_{\text{max}}^i, y_{\text{max}}^i)\}_{i=1}^{N},
\end{equation}
where $(x_\text{{min}}^i,y_\text{{min}}^i,x_\text{{max}}^i,y_\text{{max}}^i)$ represents the coordinates of the bounding box corners for the $i$-th frame. Within the AMCM module, $B$ is repeated and concatenated with the feature maps $F$ prior to the temporal transformer. The concatenated features in the $l$-th block are represented as:
\begin{equation}
    {F}_\text{{cat}}^{(l)} = \text{concat}({F}^{(l)}, B),
\end{equation}
Subsequently, the concatenated features undergo processing through the fully connected layers, followed by the temporal attention mechanism inside the AMCM. The output features are then fed into the temporal transformer. Thus, the AMCM serves as a plugin that directly connects the spatial transformer and the temporal transformer.

\subsection{Dataset}
Our dataset for transparent video generation consists of transparent images and videos. The former data is primarily collected from the Internet, comprising images mostly handcrafted by designers, all of which are licensed for research purpose. These high-quality images feature fine edges and cover a diverse range of categories, including animals, humans, and objects. We employ a thorough cleaning process, which involves removing images with entirely white alpha channels, filtering out images with resolutions below 100x100, eliminating text-only images, and manually excluding non-compliant images. Through this process, we obtain 100K transparent images.

In contrast, acquiring high-quality transparent video data presents greater challenges. Therefore, we leverage the open-source video segmentation dataset (Youtube VOS \cite{xu2018youtube}) utilizing approximately 3K videos. These are then segmented by object ID, resulting in around 10K transparent video clips, totaling approximately 150K frames. For each video, we extract the middle frame and use an image captioning model (CoCa \cite{yu2022coca}) to generate captions, creating video-text pairs for training the AMCM module.

\section{Experiments}
\subsection{Experimental Setup}
We utilize the AdamW optimizer with a learning rate of 3e-5 for both the TVAE and the AMCM training. The TVAE is trained on a single A10 GPU for 100K iterations with a batch size of 4, while the AMCM is trained on 4 A10 GPUs with a batch size of 16 for 3K iterations. All training data is resized to a resolution of 384x384, and videos are sampled at 16 frames. The loss scaling factor $\lambda$ is set to 1.

\begin{figure}[!t]
    \centering
    \includegraphics[width=0.47\textwidth]{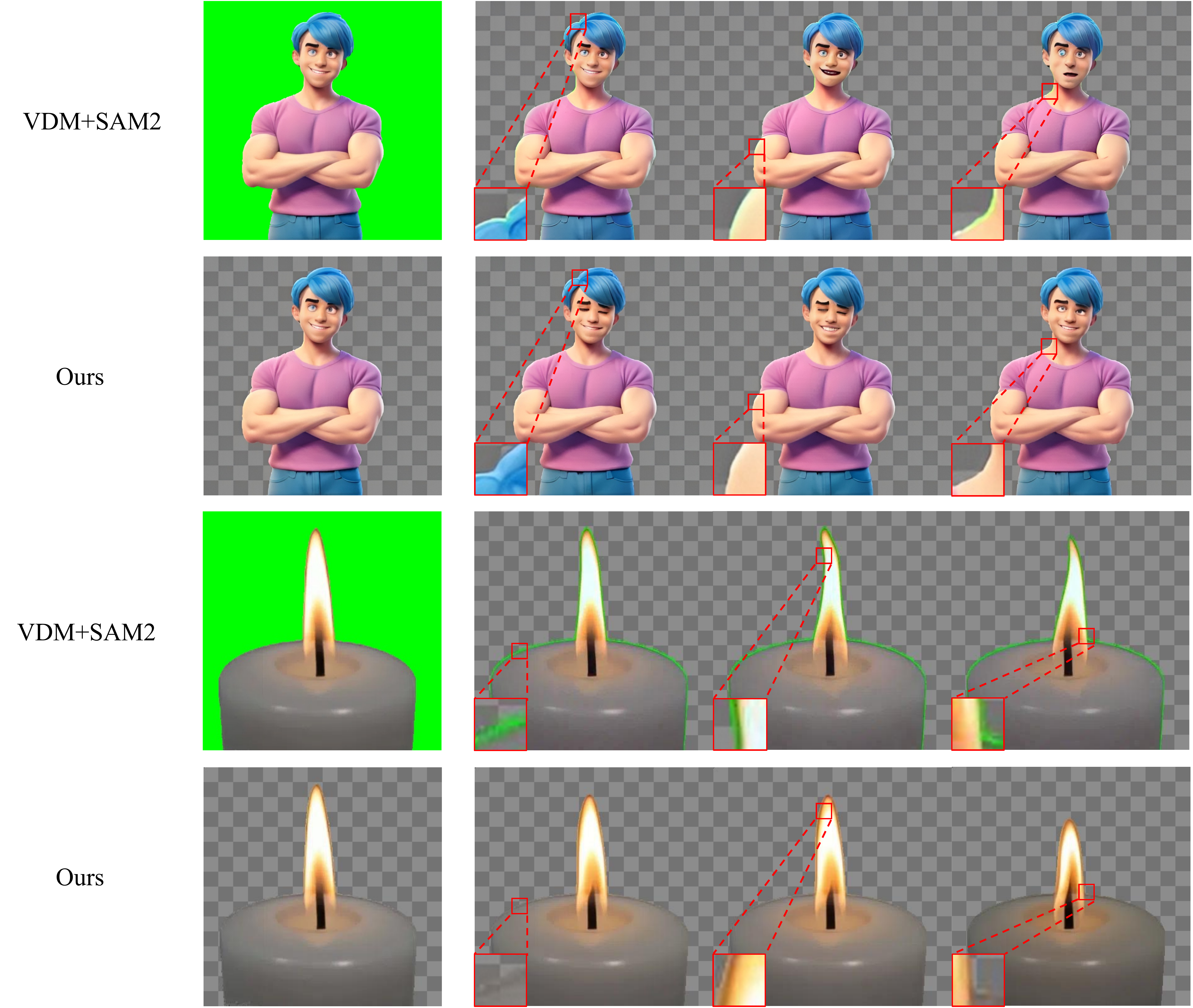}
    \caption{
        Qualitative Results. The input includes two image-text pairs, with the corresponding texts being ``a man smiling" and ``the candle is lit up".
    }
    \label{fig:qua_exp}   
\end{figure}

\begin{figure}[!t]
    \centering
    \includegraphics[width=0.47\textwidth]{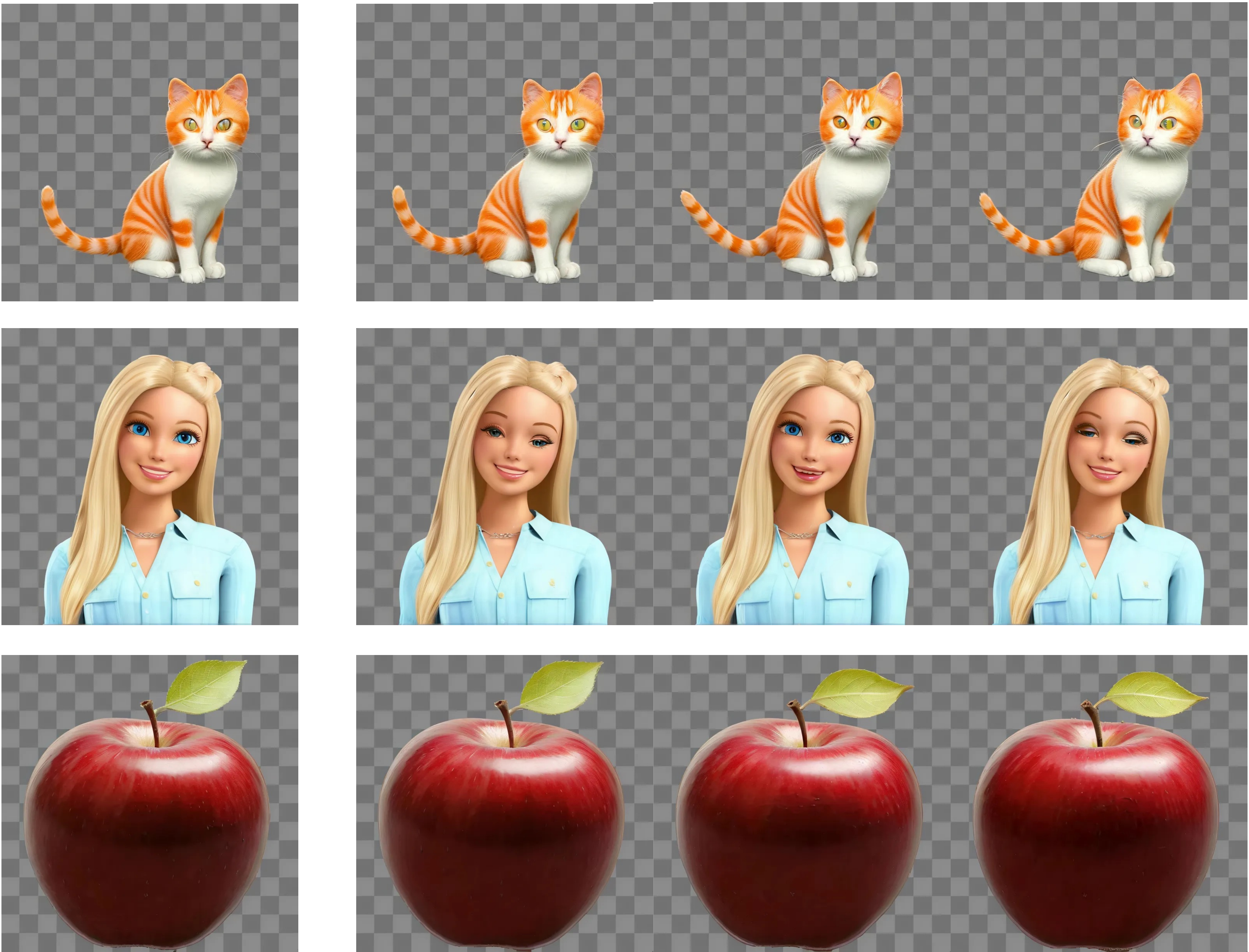}
    \caption{
        More cases generated by TransVDM, with the first column showing the conditioned images and the following columns displaying the generated outputs. The prompts are ``a cat turns head", ``Barbie blinking her eyes", and ``the apple is swaying in the wind".
    }
    \label{fig:morecase}   
\end{figure}

\subsection{Qualititive Results}
We present the results of our proposed TransVDM model compared to traditional post-processing methods (VDM+SAM2) in Fig.\ref{fig:qua_exp}. The first column showcases the conditioned images, while the subsequent columns illustrate three generated transparent frames. Our approach directly generates transparent videos from the given transparent image and text prompt. In contrast, the post-processing method requires two steps: first, it replaces the transparent regions of the input image with a specific background to produce an RGB image, which is then processed using the vanilla VDM model to generate a video, followed by removing the background. For the post-processing method, we employ a green screen technique akin to that used in the film industry, filling the transparent images with a green background to convert them into RGB format. To eliminate the background from the generated frames, common methods include segmentation (SAM2 \cite{ravi2024sam}), matting (PPMatting \cite{chen2022pp}), and chroma key. Segmentation and matting are automated algorithms, whereas chroma key necessitates manual threshold adjustments to match the green pixel values within a specified range. The results from these three methods are similar, often leaving green backgrounds around the edges of the foreground. As shown in Fig.\ref{fig:qua_exp}, we display the results of using SAM2. We observe that our method produces cleaner results, confirming the advantages of TransVDM over conventional post-processing techniques. Lastly, we show more results generated by TransVDM in Fig.\ref{fig:morecase}.

\begin{table}[!t]
\renewcommand{\arraystretch}{1.3}
\centering
\caption{Ablation Study of AMCM}
\begin{tabular}{ccc}
\hline
Method   & FVD~($\downarrow$) & LPIPS~($\downarrow$) \\ \hline
w/o AMCM & 680 & 0.222 \\
w/ AMCM   & \textbf{614} & \textbf{0.219} \\ \hline
\end{tabular}
\label{tab:study}
\end{table}

\subsection{Ablation Study}
In this section, we conduct an ablation study on a self-constructed test dataset, which consists of 100 manually collected transparent video-text pairs. We adopt two commonly used metrics in video generation: FVD and LPIPS \cite{yang2023video}. FVD evaluates the temporal quality of generated videos while LPIPS assesses the spatial visual quality of the generated results. Notably, we convert the transparent videos to RGB format for metric calculation.

As shown in Tab.\ref{tab:study} , our method demonstrates a significant improvement in terms of FVD compared to the vanilla VDM without the AMCM module, while the LPIPS remain comparable. This indicates training the AMCM module substantially enhances the performance of the generated videos, thereby confirming the effectiveness of our method.


\section{conclusion and future work}
We propose the first diffusion model for transparent video generation, which incorporates a Transparent Variational Autoencoder (TVAE), a vanilla Variational Diffusion Model (VDM), and an Alpha Motion Constraint Module (AMCM). Additionally, we have built a dataset of 250K samples to train the TVAE and AMCM. Extensive experiments demonstrate the superiority of our method. In future work, we plan to explore the possibility of manually inputting motion constraints during inference for more controllable video generation, while also enhancing our dataset regarding quality and quantity to improve the robustness of our method. Moreover, we intend to investigate more powerful VDM foundations, such as SVD \cite{blattmann2023stable} and Diffusion Transformer based models \cite{peebles2023scalable,hong2022cogvideo,zhang2024tora}, to further elevate the performance of our framework.

\clearpage
\bibliographystyle{abbrv}
\bibliography{references}
\end{document}